\documentclass[]{pasj01}
\draft
\begin{document} 
\Received{}
\Accepted{}
\title{Discovery of recombining plasma associated with the candidate supernova remnant G189.6$+$3.3 with Suzaku}

\author{
Shigeo \textsc{Yamauchi}\altaffilmark{1, $\ast$}, 
Moe \textsc{Oya}\altaffilmark{1}, 
Kumiko K. \textsc{Nobukawa}\altaffilmark{1,2}, 
and Thomas G. \textsc{Pannuti}\altaffilmark{3}
}
\altaffiltext{1}{Faculty of Science, Nara Women's University, Kitauoyanishimachi, Nara 630-8506, Japan}
\email{yamauchi@cc.nara-wu.ac.jp}
\altaffiltext{2}{Department of Physics, Kindai University, 3-4-1 Kowakae, Higashi-Osaka, Osaka 577-8502, Japan}
\altaffiltext{3}{Department of Physics, Earth Science and Space Systems Engineering, Morehead State University, 235 Martindale Drive, Morehead, KY 40351, USA}
\KeyWords{ISM: individual objects (G189.6$+$3.3) --- ISM: supernova remnants --- X-rays: ISM } 
\maketitle

\begin{abstract}
We present the results of an X-ray spectral analysis of the northeast region of the candidate supernova remnant G189.6$+$3.3 with Suzaku. 
K-shell lines from highly ionized Ne, Mg, Si, and S were detected in the spectrum for the first time. 
In addition, a radiative recombining continuum (RRC) from He-like Si was clearly seen near 2.5 keV. 
This detection of an RRC reveals for the first time that G189.6$+$3.3 possesses an X-ray-emitting recombining plasma (RP). 
The extracted X-ray spectrum in the 0.6--10.0 keV energy band is well fitted 
with a model consisting of a collisional ionization equilibrium 
plasma component (associated with the interstellar medium) and an RP component (associated with the ejecta).
The spectral feature shows that G189.6$+$3.3 is most likely to be a middle-aged SNR with an RP.
\end{abstract}

\section{Introduction}

Supernovae (SNe) release tremendous amounts of energy and are the main agents in the Universe 
for the synthesis of atoms of heavy elements. 
The supernova remnants (SNRs) help transfer the energy of SNe into the surrounding interstellar medium (ISM) and 
are the leading candidates for the acceleration of cosmic-ray particles to approximately the "knee" energy of the cosmic-ray spectrum. 
The expanding shock wave associated with SNRs sweeps up circumstellar matter and produces an X-ray-emitting plasma 
comprised of both ISM and stellar ejecta. 
One quantity that describes the plasma is the electron temperature $kT_{\rm e}$. 
This temperature, which corresponds to the temperature of the electrons of the plasma, gradually increases over time 
through Coulomb collisions between the electrons and more energetic particles in the plasma. 
Furthermore, these energetic electrons ionize neutral atoms within the plasma. 
The temperature of these ionized atoms, denoted as the ionization temperature $kT_{\rm i}$, is another quantity 
that describes the plasma and typically $kT_{\rm i}$ follows $kT_{\rm e}$. 
In general, the X-ray emitting plasmas associated with SNRs are not in a collisional ionization equilibrium (CIE) state 
where $kT_{\rm e}=kT_{\rm i}$ but instead in an ionizing plasma (IP) state where $kT_{\rm e}>kT_{\rm i}$. 
This is because the ionization dominant phase of SNRs typically lasts $>$10$^4$ yr. 
Therefore, the X-ray spectra of most young and middle-aged SNRs are well-fitted with an ionizing plasma (IP) model. 

In the last decade, strong radiative recombination continuua (RRCs) were discovered in the X-ray spectra of several Galactic SNRs, 
including IC 443 \citep{Yamaguchi2009} and W49B \citep{Ozawa2009}. 
Since the RRC originates from radiative transitions where free electrons become bound to ions, 
the strong RRC is a sign of a recombining plasma (RP), which is characterized by $kT_{\rm e}<kT_{\rm i}$. 
Examples of Galactic SNRs that possess recombining plasmas -- in addition to the SNRs mentioned above -- are 
G359.1$-$0.5 \citep{Ohnishi2011}, W28 \citep{Sawada2012}, W44 \citep{Uchida2012}, and G346.6$-$0.2 \citep{Yamauchi2013}.

Since the formation of an RP is not predicted by a standard scenario of plasma evolution after the explosion described above, 
its formation must have been driven by special circumstances in the prior history of the SNR. 
Several scenarios that may explain the formation of the RP have been proposed in the literature. 
For example, in the conduction scenario, $kT_{\rm e}$ drops below $kT_{\rm i}$ by conductive cooling by a cold cloud 
(e.g., \cite{Kawasaki2002, Matsumura2017}). 
Another scenario -- known as the rarefaction scenario -- proposes that adiabatic cooling occurs 
when the plasma breaks out from a dense medium into 
a much less dense medium (e.g., \cite{Masai1994, Yamaguchi2018}). 
Other proposed scenarios suggest that $kT_{\rm i}$ increases by either photo-ionization by an external X-ray source 
(e.g., \cite{Nakashima2013, Ono2019}) or by ionization due to low-energy cosmic rays (LECRs; e.g. \cite{Hirayama2019}). 
The precise origin of RPs associated with SNRs remains uncertain and realizing a clear understanding of 
their origin within the evolution of SNR plasmas is an outstanding unresolved issue.

The candidate SNR G189.6$+$3.3 was discovered by the ROSAT All-Sky Survey \citep{Asaoka1994}. 
The image of G189.6+3.3 revealed a ring-like X-ray morphology with a diameter of $\sim$\timeform{1.D5}. 
The center of this SNR is offset from the center of the nearby prominent SNR IC 443 by $\sim$\timeform{0.D7} (see figure 1). 
The northeast region of G189.6+3.3 has the highest surface brightness in both the soft X-ray band \citep{Asaoka1994} 
and the radio band, specifically at 327 MHz \citep{Braun1986}. 
Analysis of the extracted X-ray spectrum of G189.6$+$3.3 revealed that the X-ray emission is mainly soft 
with a characteristic temperature of 0.14 keV. 
This result indicates that G189.6$+$3.3 is approximately 10$^5$ years old and is therefore a well-evolved SNR. 
Asaoka and Aschenbach (1994) argued that G189.6$+$3.3 lies in front of a molecular cloud that is itself known to lie in front of IC 443. 
Those authors therefore estimated a distance to G189.6$+$3.3 of approximately 1.5 kpc.

After the analysis of the ROSAT observation of G189.6+3.3 by Asaoka and Aschenbach (1994), 
there has been no further detailed analysis in the literature of the X-ray properties of this SNR. 
Suzaku \citep{Mitsuda2007} observed a northeast region (the brightest region) of G189.6$+$3.3. 
We analyzed the Suzaku archival data and discovered evidence for an RP associated with G189.6$+$3.3. 
In this paper, we report results of our spectral analysis. 
The quoted errors are at the 90\% confidence level unless otherwise mentioned.

\section{Observations and Data Reduction}

The Suzaku observation of the northeast region of G189.6$+$3.3 was conducted on March 15--17 in 2015 (Obs. ID 509036010) 
with the X-ray Imaging Spectrometer (XIS: \cite{Koyama2007}). 
The observed field is shown in figure 1. 
The XIS consisted of 4 sensors: XIS sensor-1 (XIS1) is a back-side illuminated CCD (BI), while
the other three XIS sensors (XIS0, 2, and 3) are a front-side illuminated CCD (FI). 
Since XIS2 ceased to function properly in 2006, observations were made with XIS0, XIS1 and XIS3. 
A small fraction of the collecting area of XIS 0 was not used because of damage, 
possibly due to the impact of a micrometeorite on 2009 June 23.
The XIS -- which was operated in the normal clocking mode during the observation -- 
employed the spaced-row charge injection (SCI) technique to rejuvenate 
its spectral resolution by filling the charge traps with artificially injected electrons through CCD readouts.
Details concerning on the SCI technique are given in \citet{Nakajima2008} and \citet{Uchiyama2009}.

Data reduction and analysis were made using the HEAsoft version 6.25.
The XIS pulse-height data for each X-ray event were converted to 
Pulse Invariant (PI) channels using the {\tt xispi} software and the calibration database version 2018-10-10.
We screened the data using the standard criteria. 
During the observations, count rates of the non-X-ray background (NXB) of XIS 1 were systematically higher than those of the NXB data 
generated by {\tt xisnxbgen} \citep{Tawa2008}. 
Therefore, we utilized only the FI in the following analysis.  
The exposure times after applying the screening criteria are 66.1 ks and 86.2 ks for XIS0 and XIS3, respectively.

\begin{figure*}
  \begin{center}
    \includegraphics[width=10cm]{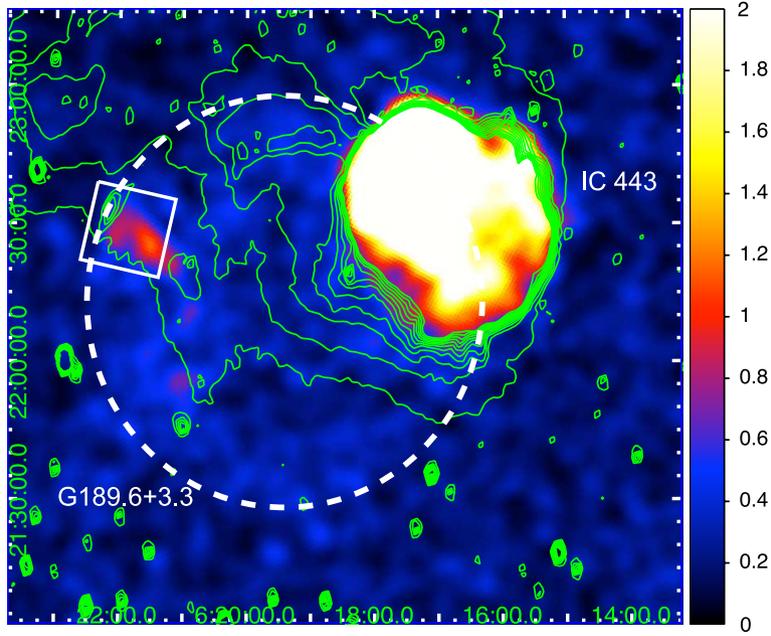} 
  \end{center}
  \caption{X-ray (ROSAT, color) and radio band (1420 MHz, processed by the Canadian Galactic Plane Survey Consortium, green contour) images 
  of G189.6$+$3.3 and IC 443 region,
  taken from the catalog of High Energy Observations of Galactic Supernova Remnants (\cite{Ferrand2012}; http://www.physics.umanitoba.ca/snr/SNRcat). 
  Only the lower intensity levels are displayed. 
  The color bar shows the intensity levels in arbitrary units. The coordinates are J2000.0.  
  The white dashed line indicates an approximate shape of G189.6+3.3, while
  the white solid line shows the Suzaku XIS FOV.
}\label{fig:img}
\end{figure*}

\section{Analysis and Results}

\subsection{Image}

Figure 2 shows X-ray images in the "soft" (0.6--3.0 keV) and "hard" (3.0--8.0 keV) energy bands of G189.6$+$3.3 using the XIS. 
Consistent with the ROSAT image and the analysis of Asaoka and Aschenbach (1994), 
the soft X-ray band image clearly shows the presence of diffuse X-ray emission. 
We compared the soft X-ray image with radio maps of this SNR \citep{Braun1986, Leahy2004} 
and found that the X-ray emission in this band is located just inside of the radio shell (see figure 1). 
On the other hand, the hard X-ray band image shows no significant emission and 
we therefore conclude that the X-ray spectrum of G189.6$+$3.3 is primarily soft.

\begin{figure*}
  \begin{center}
    \includegraphics[width=16cm]{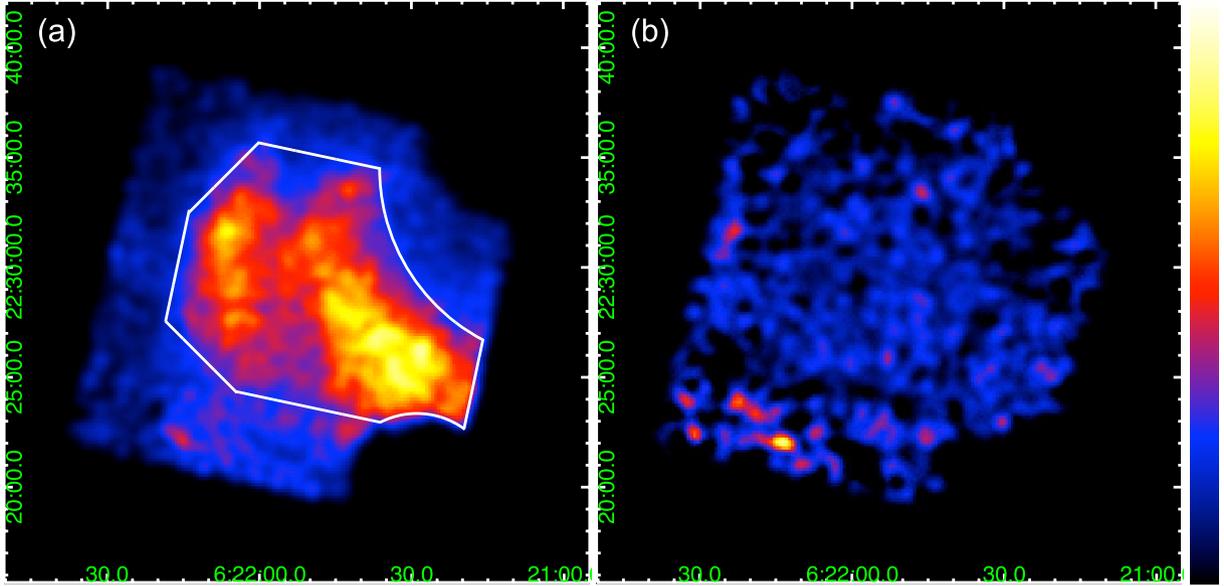} 
  \end{center}
  \caption{XIS images of the northeast region of G189.6$+$3.3: 
  (a) a soft X-ray band (0.6--3 keV) image and (b) a hard X-ray band (3--8 keV) image.
  NXB subtraction and vignetting correction were applied to both images. The coordinates are J2000.0. 
  The color scale is linear (peak to bottom).
  The white solid line shows the region from which the X-ray spectrum is extracted.
}\label{fig:img}
\end{figure*}

\subsection{Spectrum}

Based on the work of \citet{Masui2009}, we modeled the sky background spectrum, 
which consists of the Milky Way Halo (MWH), Local Hot Bubble (LHB), and Cosmic X-ray Background (CXB). 
The values of the parameters of the MWH and LHB were assumed to be the same as the values used in \citet{Hirayama2019}, 
while those of the CXB were fixed to the values in \citet{Kushino2002}. 
It is known that another diffuse source of X-ray emission known as the Galactic diffuse X-ray emission (GDXE) \citep{Koyama2018} exists along the Galactic plane. 
However, since G189.6$+$3.3 is located at the anti-center region with the Galactic latitude of $b$=\timeform{3.D3}, 
where the contamination of the GDXE is small \citep{Uchiyama2013,Yamauchi2016}, 
we ignored this component of diffuse emission in our spectral analysis. 

The source spectrum of G189.6$+$3.3 was extracted from the region indicated in figure 2. 
The NXB was estimated using {\tt xisnxbgen} \citep{Tawa2008} and was subtracted from the source spectrum. 
Figure 3 shows the NXB-subtracted spectrum of G189.6$+$3.3: 
the contributions of the sky background from the different components (MWH$+$LHB$+$CXB) are indicated with a solid gray line.

\begin{figure*}
  \begin{center}
  \includegraphics[width=8cm]{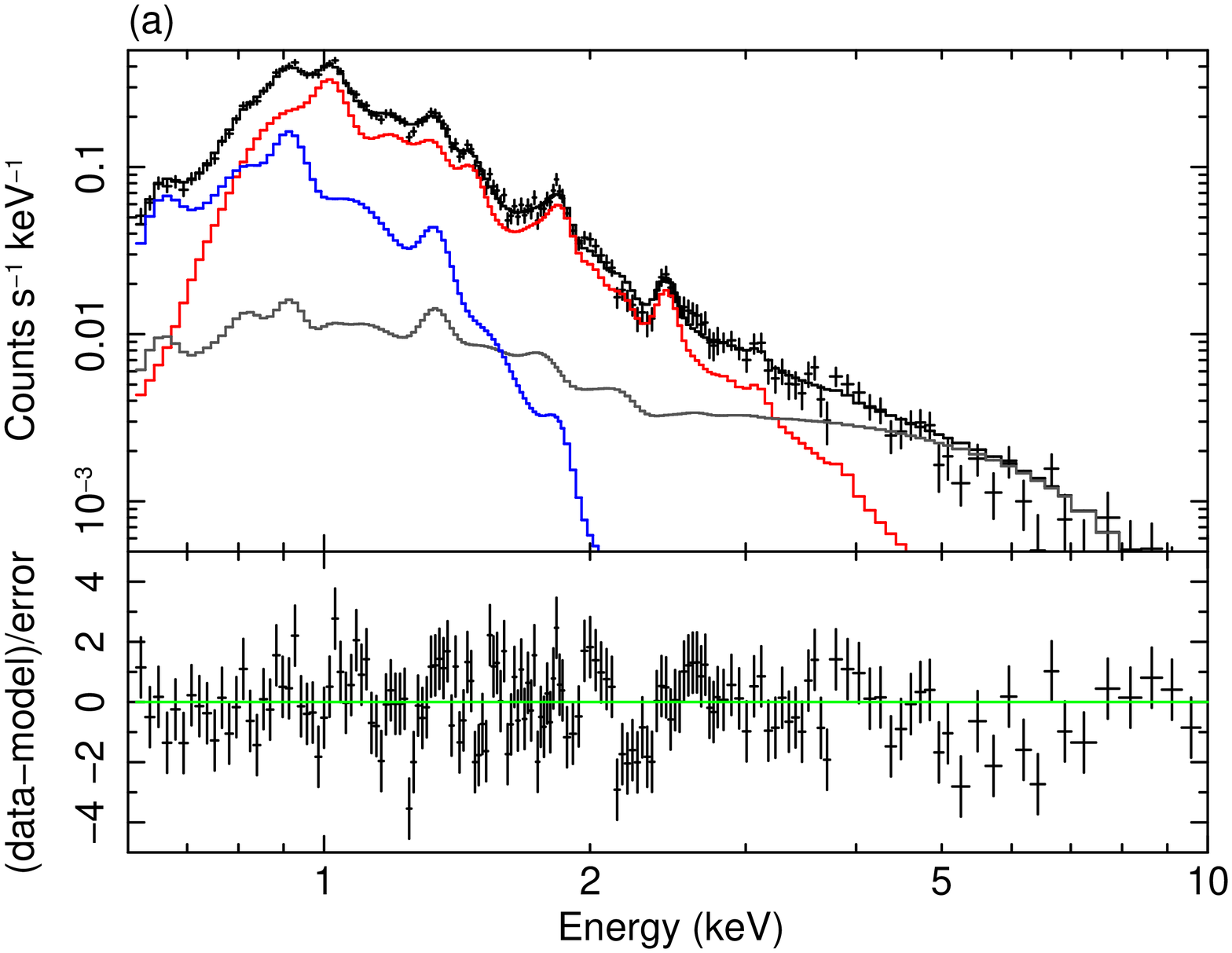} 
  \includegraphics[width=8cm]{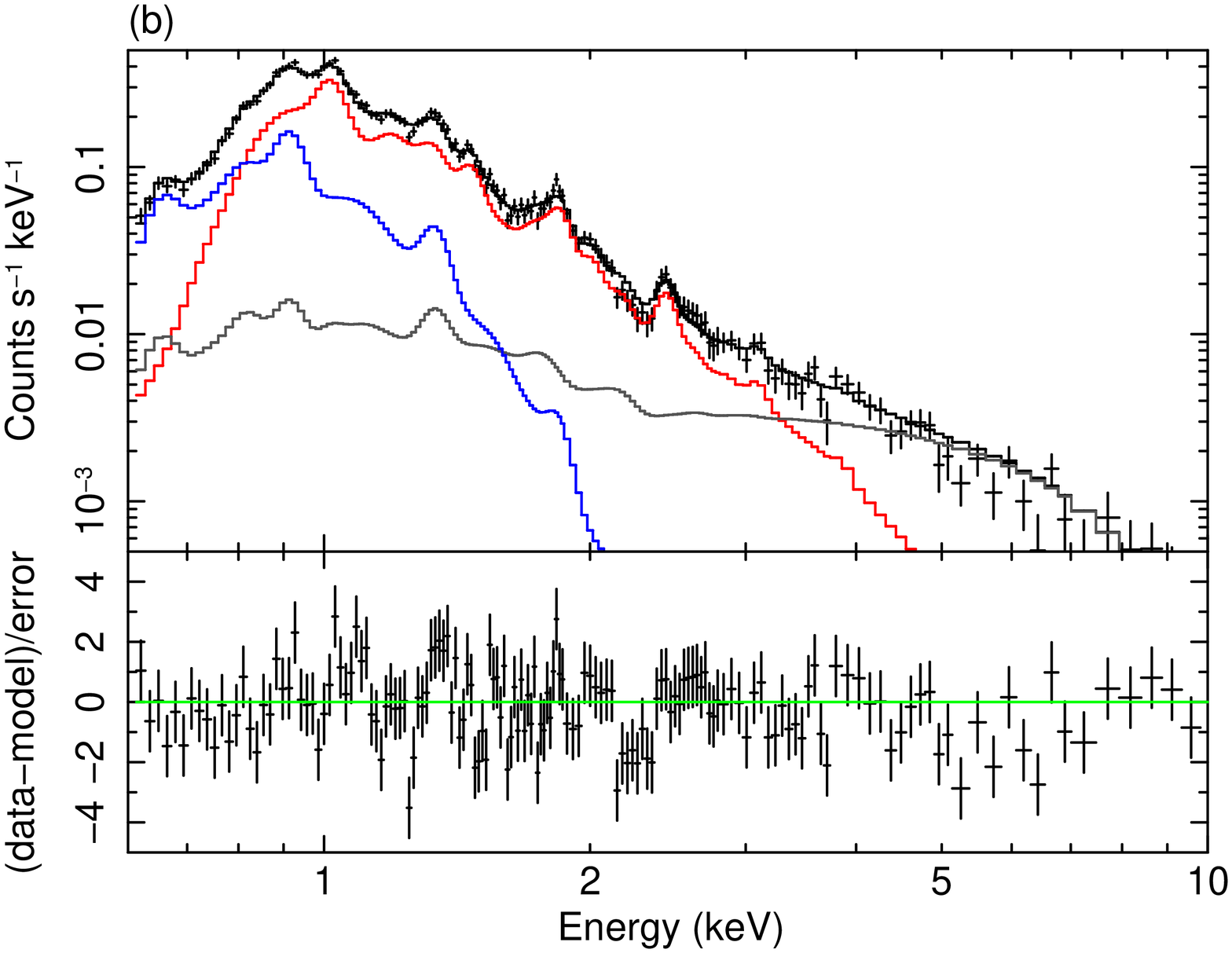} 
  \includegraphics[width=8cm]{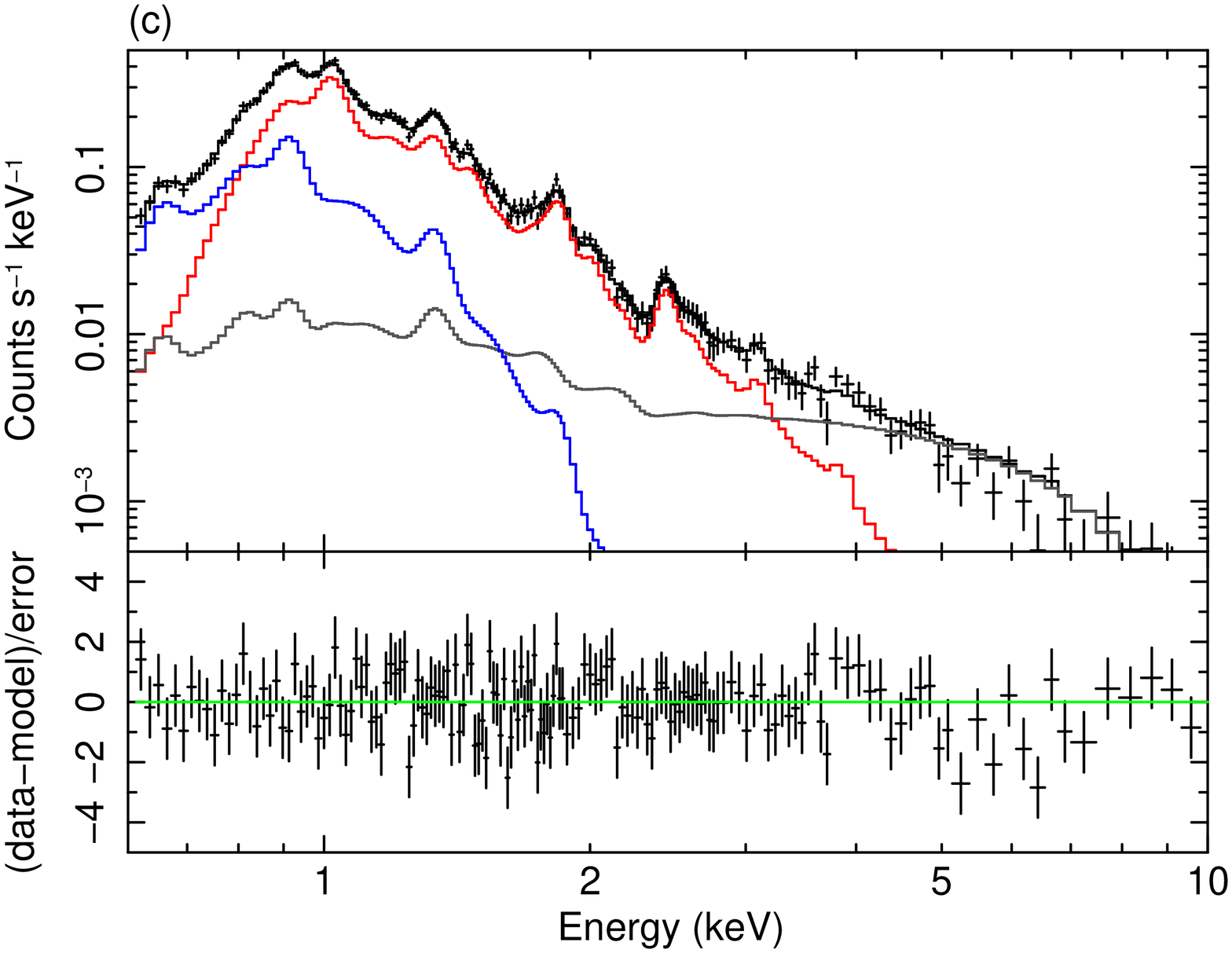} 
  \includegraphics[width=8cm]{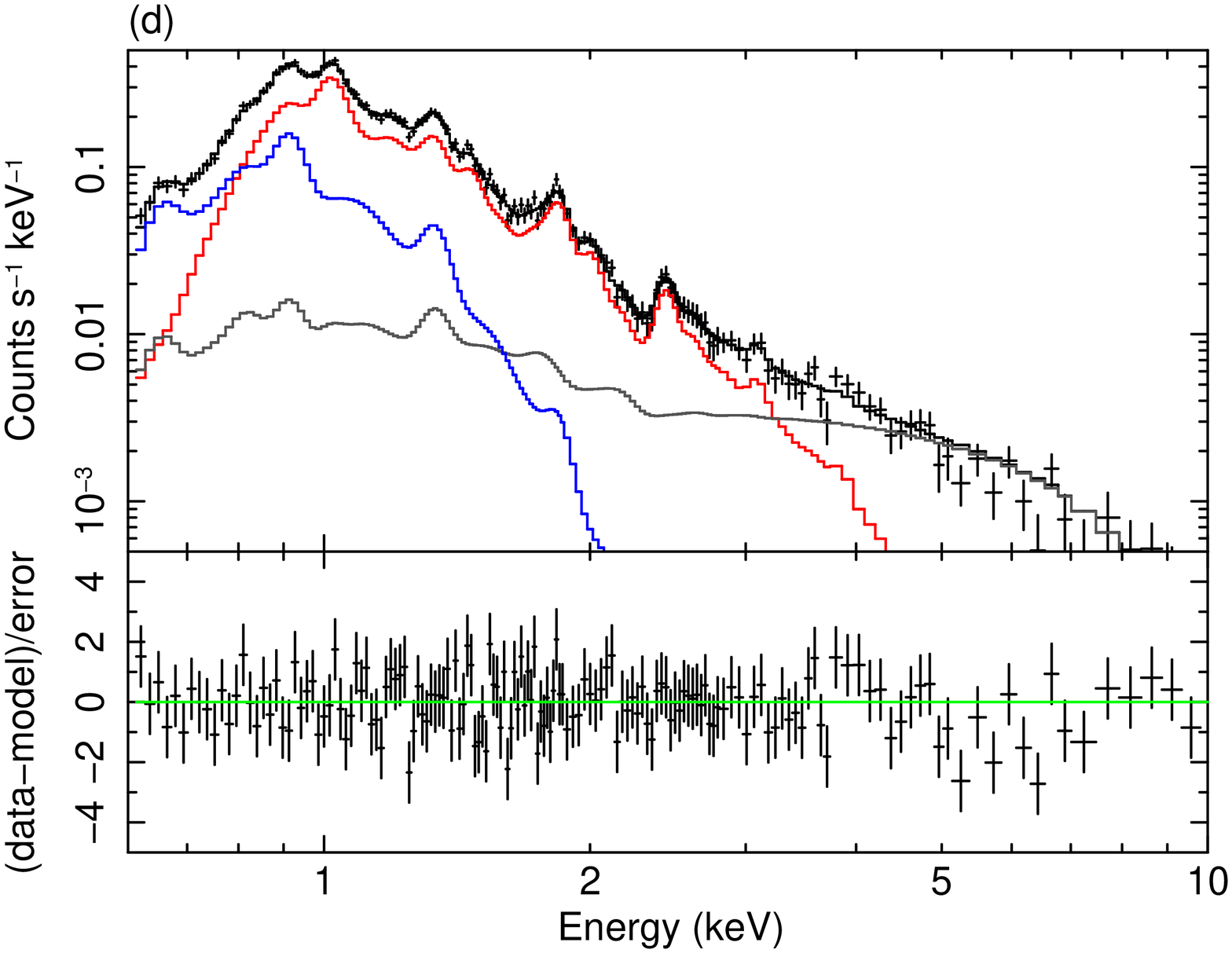} 
  \end{center}
  \caption{XIS spectrum of the east region (upper panel) and residuals from the best-fit model (lower panel), (a) model A, (b) model B, (c) model C, and (d) model D. 
  Errors of the data points are at the 1$\sigma$ level. 
  The blue, red, and gray solid lines show emission from ISM, ejecta, and the sky background (MWH$+$LHB$+$CXB) components, respectively. 
}\label{fig:img}
\end{figure*}

To fit the extracted source spectrum, we first applied a CIE model (corresponding to the {\tt vapec} model in XSPEC) 
combined with a low-energy absorption model. 
The cross section of the photoelectric absorption and the abundance tables were taken from Balucinska-Church and McCammon (1992) and 
Anders and Grevesse (1989), respectively. 
In the spectral fitting, we corrected the energy scale using a linear function.  
Residuals were seen in the spectrum in the 0.6--1.0 keV band: to address these, 
we added another CIE model with solar abundances to represent emission from a shocked ISM component.
We denote this first combined model for fitting as Model A and 
note that Model A failed to provide a statistically-acceptable fit to the source spectrum ($\chi^2$/d.o.f.=224.6/151=1.49).
We also found systematic residuals around 2.5 keV: the presence of these residuals --  
combined with positive residuals at the energy of the Si Ly$\alpha$ line -- 
suggested the presence of an RRC from He-like Si (see figure 3a).

The implied presence of an RRC motivated us to apply an RP model (corresponding to the {\tt vrnei} model in XSPEC) 
for the ejecta component of the observed X-ray emission.  
Similar to our results with the CIE model, 
we again found residuals in the 0.6--1.0 keV energy range, and added the shocked ISM component. 
We denote this combined CIE$+$RP model as Model B and note that this combined model also failed to yield 
a statistically-acceptable fit to the spectrum  ($\chi^2$/d.o.f.=235.4/150=1.57). 
Inspection of the fit obtained with this model revealed that the Mg He$\alpha$ line and the RRC structure remain in the fit residuals 
(see figure 3b). 
These suggest that $kT_{\rm e}$ of the lighter elements differs from $kT_{\rm e}$ of the heavier elements. 
As our next step, we applied a two-electron temperature RP model (denoted as Model C). 
Specifically, the first RP component (denoted as RP1) consists of the elements H through Si, 
while the second component (denoted as RP2) consists of S through Ni. 
While the electron temperatures of RP1 and RP2 were varied as free parameters independent of each other, 
other parameters, specifically the recombination timescale, the initial temperature $kT_0$, and the normalizations, were tied together. 
We found at last that Model C gave an acceptable fit to the source spectrum with $\chi^2$/d.o.f.=151.0/148=1.02 (see table 1 and figure 3c). 

In the case of Model C, the plasma is assumed to have the same $kT_{\rm i}$ for all the elements (which is consistent with CIE)
at the epoch where the plasma made a transition into an RP. 
However, we note that attaining CIE is not an inevitable condition for plasmas associated with SNRs. 
For example, in analyzing Suzaku data for W28, Sawada and Koyama (2012) examined the time evolution of RPs 
with the same $kT_0$ for all elements (again consistent with CIE) and a different $kT_0$ for each element,  
and showed that the spectrum could be fit adequately in both cases. 
Furthermore, \citet{Hirayama2019} applied a multi-$kT_{\rm 0}$ RP model, where each element has its own value for $kT_0$, 
to the high-quality spectrum of IC 443, and showed the spectrum is represented by the model. 
We followed the examples by these authors and attempted to fit the source spectrum of G189.6$+$3.3 
with a multi-$kT_0$ RP model, 
which we denote as Model D. 
We obtained a statistically-acceptable fit with this model as well ($\chi^2$/d.o.f.=146.1/144=1.01). 
The improvement from the Model C fit is not statistically significant. 
The best-fitting parameters of this model are listed in table 1 and the model itself is plotted in figure 3d.

\begin{table*}
\small
\caption{The best-fit parameters of CIE (Model A), single-RP (Model B), two-temperature RP (Model C), and multi-$kT_{\rm 0}$ RP (Model D).}
 \begin{center}
 \begin{tabular}{llccccccc} \hline
Component & Parameter & \multicolumn{7}{c}{Values} \\
& & Model A & \multicolumn{2}{c}{Model B} & \multicolumn{2}{c}{Model C} & \multicolumn{2}{c}{Model D} 
\\ \hline
Absorption& 
$N_{\rm H}^{\ast}$ & 
6.0$\pm$0.4 & \multicolumn{2}{c}{5.9$\pm$0.4} &  \multicolumn{2}{c}{5.9$\pm$0.4}  &  \multicolumn{2}{c}{6.2$\pm$0.7}    \\ \hline
ISM\ (CIE) & $kT_{\rm e}^{\dag}$ & 
0.18$\pm$0.01 &\multicolumn{2}{c}{0.18$\pm$0.01} &  \multicolumn{2}{c}{0.18$\pm$0.02}  &  \multicolumn{2}{c}{0.18$\pm$0.02}   \\
& Normalization$^{\ddag}$ & 
0.038$\pm$0.010 & \multicolumn{2}{c}{0.035$\pm$0.009} &  \multicolumn{2}{c}{0.031$\pm$0.010}  &  \multicolumn{2}{c}{0.038$\pm$0.014}    \\ \hline
Ejecta\ (RP1) & $kT_{\rm e}^{\dag}$ & 
0.79$\pm$0.04 & \multicolumn{2}{c}{0.78$\pm$0.05} &  \multicolumn{2}{c}{0.45$\pm$0.04}  &  \multicolumn{2}{c}{0.46$\pm$0.07}    \\
& Normalization$^{\ddag}$ & 
0.0030$\pm$0.0003 & \multicolumn{2}{c}{0.0028$\pm$0.003} &  \multicolumn{2}{c}{00030$\pm$0.004}  &  \multicolumn{2}{c}{0.0032$\pm$0.0011}   \\      
& $n_{\rm e}t^{\S}$ & --- 
& \multicolumn{2}{c}{11$^{+7}_{-3}$} &  \multicolumn{2}{c}{6.0$\pm$0.7}  &  \multicolumn{2}{c}{1.8$^{+4.5}_{-1.2}$}   \\ 
&   & $Ab^{\Vert}$ & $kT_{\rm 0}^{\dag}$ & $Ab^{\Vert}$  & $kT_{\rm 0}^{\dag}$ & $Ab^{\Vert}$ & $kT_{\rm 0}^{\dag}$ & $Ab^{\Vert}$  \\  
&    H=He=C=N=O 	&1 (fixed)		&  (link to Ne)   		& 1 (fixed) 		&  (link to Ne)  		& 1 (fixed) 		&  (link to Ne) 			& 1 (fixed)         \\
&    Ne 			&6.0$\pm$1.0	& 3.0 (fixed)  		& 6.7$\pm$1.1 		& 3.0$^{+2.2}_{-0.9}$ 	& 6.1$\pm$1.9 	& 0.50$^{+0.71}_{-0.08}$	& 5.1$^{+3.1}_{-2.2}$         \\
&    Mg 			&1.2$\pm$0.2	& (link to Ne)  		& 1.3$\pm$0.3 		& (link to Ne) 		& 3.0$\pm$0.6  	& 0.79 ($>$0.69) 		& 2.8$^{+1.2}_{-0.8}$		 \\
&    Si 			&0.8$\pm$0.2	& (link to Ne)  		& 0.9$\pm$0.2 		& (link to Ne) 		& 3.2$\pm$0.7 		& 1.2 ($>$1.0) 			& 3.0$\pm$1.0		 \\
&    S=Ar=Ca		&1.0$\pm$0.3	& (link to Ne)  		& 1.1$\pm$0.3		& --- 		&0 (fixed)			& ---				& 0 (fixed)	 \\
&    Fe=Ni 		&1 (fixed)		& (link to Ne)  		& 1 (fixed) 		& ---		& 0 (fixed)			& ---  			& 0 (fixed) \\
\hline
Ejecta\ (RP2) & $kT_{\rm e}^{\dag}$ & --- & \multicolumn{2}{c}{---} &  \multicolumn{2}{c}{0.73$\pm$0.05}  &  \multicolumn{2}{c}{0.67$\pm$0.13}    \\
& Normalization$^{\ddag}$ & --- & \multicolumn{2}{c}{---} &  \multicolumn{2}{c}{(link to RP1)}  &  \multicolumn{2}{c}{(link to RP1)}   \\      
& $n_{\rm e}t^{\S}$ & ---& \multicolumn{2}{c}{---} &  \multicolumn{2}{c}{(link to RP1)}  &  \multicolumn{2}{c}{(link to RP1)}   \\ 
&    H--Si			&---			& ---  			& --- 				& --- 				& 0 (fixed) 		& --- 						& 0 (fixed)		 \\
&    S=Ar=Ca		&---			& ---  			& --- 				& (link to Ne) 		& 1.8$\pm$0.5 		& 1.5$^{+2.2}_{-0.3}$ 		& 1.9$\pm$0.6		 \\
&    Fe=Ni 		&---			& ---  			& --- 				& (link to Ne) 		& 1 (fixed)		&1.1$^{+3.4}_{-0.3}$ 	& 1 (fixed)	 \\ \hline
& $\chi^2$/d.o.f. & 224.6/151=1.49& \multicolumn{2}{c}{235.4/150=1.57} &  \multicolumn{2}{c}{151.0/148=1.02}  &  \multicolumn{2}{c}{146.1/144=1.01}      \\
     \hline
   \end{tabular}
   \end{center}
$^{\ast}$ The unit is $\times$10$^{21}$ cm$^{-2}$. \\
$^{\dag}$ Units are keV. $kT_{\rm e}$ is an electron temperature at the present time and $kT_{\rm 0}$ is an initial ionization temperature at $n_{\rm e}t$=0. \\
$^{\ddag}$ Defined as 10$^{-14}$ $\times$$\int n_{\rm H} n_{\rm e} dV$ / (4$\pi D^2$) (cm$^{-5}$),
where $D$ is the distance (cm), $n_{\rm H}$ is the hydrogen density (cm$^{-3}$), $n_{\rm e}$ is the electron density (cm$^{-3}$), and $V$ is the volume (cm$^3$). \\
$^{\S}$ Recombination time scale, where $n_{\rm e}$ is the electron density (cm$^{-3}$) and $t$ is the elapsed time (s). The unit is $\times$10$^{11}$ cm$^{-3}$ s. \\
$^{\Vert}$ Relative to the solar values in Anders and Grevesse (1989). \\
\normalsize
\end{table*}

\section{Discussion}

Our work presented here confirmed that G189.6$+$3.3 features an extended X-ray emission 
that is located just inside a shell-like feature observed in the radio band. 
The results of our spectral analysis revealed that the X-ray emission from stellar ejecta associated with this SNR 
is best-represented by an optically-thin thermal plasma model with an electron temperature $kT_{\rm e}$$\sim$0.4--0.7 keV. 
In addition, an RRC from He-like Si is clearly detected around 2.5 keV: the presence of this RRC is evidence 
that the X-ray emitting plasma associated with G189.6$+$3.3 is an RP. 
The spectral features of this SNR resemble other middle-age SNRs that also possess an RP 
(e.g., \cite{Yamaguchi2009, Ohnishi2011, Sawada2012, Uchida2012, Yamauchi2013}). 
Therefore, G189.6$+$3.3 is also most likely to be a middle-aged SNR with an RP.

We obtained statistically-acceptable fits to the extracted X-ray spectra of G189.6$+$3.3 with Model C 
(a two-temperature RP model) and Model D (a multi-$kT_0$ RP model). 
In the case of the fit with Model C, the plasma is assumed to be in the CIE state at the epoch of transitioning into an RP. 
Note that the fitted initial temperature of the plasma with Model C is $\sim$3 keV. 
A CIE plasma with this temperature is not typically seen in in SNRs evolving in relatively low-density environments, 
but in the case of a supernova explosion that occurs in a high-density environment, a high temperature CIE plasma 
may be produced (e.g., \cite{Katsuda2016}). 
In contrast, Model D represents an RP which is not initially in CIE. 
To explore this result further, we simulated IP spectra assuming various values of ionization timescales and $kT_{\rm e}$, 
and fitted these spectra with a multi-$kT_{\rm i}$ model which has a different value for $kT_{\rm i}$ for each element. 
With this approach, we found that an IP spectrum with $kT_{\rm e}$$\sim$ 2 keV and an ionization timescale of 
$\sim$2$\times$10$^{11}$ cm$^{-3}$ s is approximated by the multi-$kT_{\rm i}$ model in which $kT_{\rm e}$ is 
$\sim$2 keV and $kT_{\rm i}$ of each element is consistent with $kT_0$ of Model D. 
This result indicates that the initial condition described by the fitted parameters of Model D can approximate the 
non-equilibrium ionization state. 
Furthermore, the $kT_0$ values of Model D appear to increase with increasing atomic number over the range of elements 
from Ne to Ca. 
This result parallels those of the different $kT_0$ case attained by an IP as shown by Sawada and Koyama (2012). 
Therefore, an acceptable fit to the spectra using Model D implies that a scenario 
where the plasma transitions from an IP to an RP is possible. 
We note that Model D assumes the ion population of a CIE plasma with $kT_{\rm 0}$ for each element, which is actually different from that of a real IP. 
To investigate this scenario, a plasma model taking into account the evolution from an IP at the initial epoch to an RP is needed.  

As described previously, different scenarios to produce an RP, such as the thermal conduction scenario, 
the adiabatic expansion scenario, 
the scenario of photo-ionization by an external source and the scenario 
where ionization is accomplished by LECRs, have been presented in the literature. 
As applied to G189.6$+$3.3, we rule out the scenario of photo-ionization by an external source 
because no such bright sources have been found near this SNR so far. 
The remaining three scenarios cannot be ruled out at this time: both the thermal conduction and adiabatic expansion scenarios 
suggest that $kT_{\rm e}$ dropped significantly and therefore the plasma transitioned to an RP from either an IP or a state of CIE. 
To investigate either of these scenarios further, it is crucial to determine if the ISM that surrounds G189.6$+$3.3 is particularly 
dense or not. 
Regarding the scenario where ionization is accomplished by LECRs, it is important to note that the prominent X-ray emission 
from this SNR is located near the shell-like structure detected in the radio band (see figure 1). 
This result suggests that LECRs accelerated at the shell may be ionizing atoms and establishing the condition 
where $kT_{\rm e}$$<$$kT_{\rm i}$. 
For this scenario, however, more evidence is needed to indicate that particle acceleration is indeed occurring at this location. 
Since information about the physical conditions of the ISM surrounding G189.6+3.3 is limited, 
at this point we cannot further distinguish between any of these three scenarios for the origin of the RP.

The parameters of the best-fits with Model C and Model D indicate that ions of lighter elements are described 
with lower electron temperatures. 
A similar relation between the atomic numbers of elements and their temperatures was found in an X-ray study of 
the Galactic SNR G272.2$-$3.2 \citep{Kamitsukasa2016}. 
This result may indicate that the plasma associated with G189.6$+$3.3 has a stratified structure with lighter elements 
located in outer layers of the plasma. 
In such a situation, the outer layers of the plasma would cool as the SNR expands. 

Assuming the distance to be 1.5 kpc and the light-of-sight length of the thin thermal plasma to be 4.4 pc 
(corresponding to an angular extent of 10 arcminutes), 
we calculated several physical parameters of the X-ray emitting plasma associated with this SNR. 
Adopting a filling factor of 1, the volume of the plasma of $\sim$3$\times$10$^{57}$ cm$^{-3}$ (=12$'$$\times$10$'$$\times$10$'$), 
and $n_{\rm e}$=1.2$n_{\rm H}$, where $n_{\rm e}$ and $n_{\rm H}$ are the electron and hydrogen densities, respectively, and 
applying the best-fit parameters, we obtained values for the mean $n_{\rm H}$, the gas mass, and the thermal energy of 
the ISM component to be $\sim$0.5 cm$^{-3}$, $\sim$2$M_{\odot}$, and $\sim$1.5$\times$10$^{48}$ erg, respectively. 
Similarly, for the mean $n_{\rm H}$, gas mass, and thermal energy of the ejecta component, we derived values of 
$\sim$0.15 cm$^{-3}$, $\sim$0.5$M_{\odot}$, and $\sim$10$^{48}$ erg, respectively. 
These calculated values for the mass and the thermal energy of the X-ray-emitting plasma associated with 
G189.6$+$3.3 indicate that the plasma only comprises a small portion of the total mass and total thermal energy of 
the whole SNR, when compared to other Galactic SNRs. 

While our work has revealed the presence of an RP associated with G189.6$+$3.3, 
many properties of this intriguing SNR remain unknown. 
We strongly encourage new multi-wavelength observations of this source to probe its properties in more detail. 

\begin{ack}
We would like to express our thanks to all of the Suzaku team. 
The authors wish to thank the referee for constructive comments that improved the manuscript. 
This work was supported by the Japan Society for the Promotion of Science (JSPS) KAKENHI Grant Numbers JP16J00548 (KKN). 
\end{ack}



\begin{thebibliography}{}
\bibitem[Anders \& Grevesse(1989)]{Anders1989}
   Anders, E., \& Grevesse, N. 1989, Geochim. Cosmochim. Acta, 53, 197
\bibitem[Asaoka \& Aschenbach(1994)]{Asaoka1994}
   Asaoka, I., \& Aschenbach, B. 1994, \aap, 274, 573
\bibitem[Balucinska-Church \& McCammon(1992)]{bcmc1992} 
   Balucinska-Church, M., \& McCammon, D. 1992, \apj, 400, 699
\bibitem[Braun \& Strom(1986)]{Braun1986}
   Braun, R., \& Strom, R. G. 1986, \aap, 164, 193
\bibitem[Ferrand \& Safi-Harb(2012)]{Ferrand2012}
   Ferrand, G., \& Safi-Harb, S. 2012, Adv. Space Res., 49, 1313
\bibitem[Hirayama et al.(2019)]{Hirayama2019}
   Hirayama, A.,  Yamauchi, S., Nobukawa, K. K., Nobukawa, M., \& Koyama, K. 2019, \pasj, 71, 37
\bibitem[Kamitsukasa et al.(2016)]{Kamitsukasa2016}
   Kamitsukasa, F., Koyama, K., Nakajima, H., Hayashida, K., Mori, K., Katsuda, S., Uchida, H., \& Tsunemi, H. 2016, \pasj, 68, S7
\bibitem[Katsuda et al.(2016)]{Katsuda2016}
   Katsuda, S., et al. 2016, \apj, 832, 194
\bibitem[Kawasaki et al.(2002)]{Kawasaki2002}
  Kawasaki, M. T., Ozaki, M., Nagase, F., Masai, K., Ishida, M., \& Petre, R. 2002, \apj, 572, 897
\bibitem[Koyama(2018)]{Koyama2018}
   Koyama, K. 2018, \pasj, 70, 1
\bibitem[Koyama et al.(2007)]{Koyama2007}
   Koyama, K., et al.\ 2007, \pasj, 59, S23
\bibitem[Kushino et al.(2002)]{Kushino2002}
   Kushino, A., Ishisaki, Y., Morita, U., Yamasaki, N. Y., Ishida, M., Ohashi, T., \& Ueda, Y. 2002, \pasj, 54, 327
\bibitem[Leahy(2004)]{Leahy2004}
   Leahy, D.A. 2004, \aj, 127, 2277
\bibitem[Masai(1994)]{Masai1994}
   Masai, K. 1994, \apj, 437, 770
\bibitem[Masui et al.(2009)]{Masui2009}
   Masui, K., Mitsuda, K., Yamasaki, N. Y., Takei, Y., Kimura, S., Yoshino, T., \& McCammon, D. 2009, \pasj, 61, S115
\bibitem[Matsumura et al.(2017)]{Matsumura2017}
   Matsumura, H., Tanaka, T., Uchida, H., Okon, H., \& Tsuru, T. G. 2017, \apj, 851, 73
\bibitem[Mitsuda et al.(2007)]{Mitsuda2007}
   Mitsuda, K., et al.\ 2007, \pasj, 59, S1
\bibitem[Nakajima et al.(2008)]{Nakajima2008}
   Nakajima, H., et al. 2008, \pasj, 60, S1
\bibitem[Nakashima et al.(2013)]{Nakashima2013}
   Nakashima, S., Nobukawa, M., Uchida, H., Tanaka, T., Tsuru, T. G., Koyama, K., Murakami, H., \& Uchiyama, H. 2013, \apj, 773, 20
\bibitem[Ohnishi et al.(2011)]{Ohnishi2011}
   Ohnishi, T., Koyama, K., Tsuru, T. G., Masai, K., Yamaguchi, H., \& Ozawa, M. 2011, \pasj, 63, 527
\bibitem[Ono et al.(2019)]{Ono2019}
   Ono, A., Uchiyama, H., Yamauchi, S., Nobukawa, M., Nobukawa, K. K., \& Koyama, K. 2019, \pasj, 71, 52
\bibitem[Ozawa et al.(2009)]{Ozawa2009}
   Ozawa, M., Koyama, K., Yamaguchi, H., Masai, K., \& Tamagawa, T. 2009, \apj, 706, L71
\bibitem[Sawada \& Koyama(2012)]{Sawada2012}
   Sawada, M., \& Koyama, K., 2012, \pasj, 64, 81
\bibitem[Tawa et al.(2008)]{Tawa2008}
   Tawa, N., et al. 2008, \pasj, 60, S11
\bibitem[Uchida et al.(2012)]{Uchida2012}
   Uchida, H., et al. 2012, \pasj, 64, 141
\bibitem[Uchiyama et al.(2009)]{Uchiyama2009}
   Uchiyama, H., et al. 2009, \pasj, 61, S9
\bibitem[Uchiyama et al.(2013)]{Uchiyama2013}
   Uchiyama, H., Nobukawa, M., Tsuru, T. G., \& Koyama, K. 2013, \pasj, 65, 19
\bibitem[Yamaguchi et al.(2009)]{Yamaguchi2009}
   Yamaguchi, H., Ozawa, M., Koyama, K., Masai, K., Hiraga, J. S., Ozaki, M., \& Yonetoku, D. 2009, \apj, 705, L6
\bibitem[Yamaguchi et al.(2018)]{Yamaguchi2018}
   Yamaguchi, H., et al. 2018, \apj, 868, L35
\bibitem[Yamauchi et al.(2013)]{Yamauchi2013}
   Yamauchi, S., Nobukawa, M., Koyama, K., \& Yonemori, M. 2013, \pasj, 65, 6
\bibitem[Yamauchi et al.(2016)]{Yamauchi2016}
    Yamauchi, S., Nobukawa, K. K., Nobukawa, M., Uchiyama, H., \& Koyama, K. 2016, \pasj, 68, 59
\end{thebibliography}
\end{document}